%% file: main.tex
\documentclass[a4paper,onecolumn,11pt]{quantumarticle}
\pdfoutput=1
\usepackage[english]{babel}
\usepackage[utf8]{inputenc}
\usepackage[T1]{fontenc}
\usepackage{amsmath}
\usepackage{hyperref}

\usepackage{tikz}

\usepackage{graphicx}% Include figure files
\usepackage{dcolumn}% Align table columns on decimal point
\usepackage{bm}% bold math
\usepackage{amsmath}
\usepackage{amsfonts}
\usepackage{subcaption}                 % needed for using subfigures, supersedes subfig
\usepackage{braket}
\usepackage{listings}
\usepackage{xcolor}
\usepackage{booktabs} % for \toprule etc.
\usepackage{array}    % for \arraybackslash
\usepackage{cleveref}                   % must come after hyperref, see https://ctan.org/pkg/cleveref and https://www-uxsup.csx.cam.ac.uk/pub/tex-archive/macros/latex/contrib/cleveref/cleveref.pdf
\usepackage{adjustbox}
\usepackage{caption}

\usepackage[normalem]{ulem}   % for striking through text in draft

\graphicspath{{figures/}}

\begin{document}

\title{Approximate Quantum Linear Solvers for Hybrid CFD: End-to-End Analysis with a Chebyshev-LCU Approach}

\author[1]{Tomer Goldfriend}  \email{tomer@classiq.io}%
\author[2]{Leigh Lapworth}    \email{leigh.lapworth@rolls-royce.com}%
\author[1]{Nadav Yoran}       \email{nadav@classiq.io}%
\author[1]{Amir Naveh}       \email{amir@classiq.io}%

\affil[1]{Classiq Technologies.3 Daniel Frisch Street, Tel Aviv-Yafo, 6473104, Israel.}
\affil[2]{Rolls-Royce plc, P.O. Box 31, Derby, DE24 8BJ, UK}

%\date{\today}

\begin{abstract}
Quantum linear solvers are well studied as standalone quantum algorithms; however, in hybrid classical-quantum routines, their practical value must be evaluated at the level of the full non-linear
application. A central issue is whether the approximation error of the quantum linear solver remains controlled once embedded in a full iterative workflow. We study this question in the context of a hybrid computational fluid dynamics (CFD) scheme. Through numerical simulations, we analyze how an approximate quantum linear solver affects the convergence of the overall CFD iteration. We show that convergence can be preserved for a non-exact quantum solver with only a moderate overhead in iteration count, provided that the high-frequency components of the linear system are resolved with sufficient accuracy. In addition, we develop an approximate qubitization-based solver (Cheb-LCU) that can reduce quantum resource requirements relative to a Quantum Singular Value Transformation (QSVT)-based solver while inducing only a small loss in convergence performance. This claim is demonstrated through explicit implementation and compilation of the quantum algorithms and by examining their impact on the convergence of the full CFD scheme. We find that our approximate approach reduces the required number of single-qubit rotations by over an order of magnitude relative to the QSVT-based solver, while requiring only a modest increase in CFD iteration count.
\end{abstract}

\maketitle

\section{Introduction}

The use of quantum algorithms to accelerate the solution of partial differential equations has attracted growing interest for applications in engineering and design, particularly in computational fluid dynamics (CFD)~\cite{Succi_etal2023} (and references therein). Most quantum algorithmic approaches to CFD are built on quantum linear solvers (QLSs), such as the Harrow-Hassidim-Lloyd (HHL) method~\cite{HHL} or polynomial-approximation-based techniques~\cite{Childs_etal2017}, which offer the potential for computational advantages over classical linear solvers. In many cases, quantum CFD applications are formulated within hybrid quantum-classical workflows, where the QLS is applied iteratively, either to advance the spatial solution in time or as part of an iterative correction scheme.

In such workflows, the practical value of a QLS cannot be assessed from asymptotic complexity alone, but must be evaluated in the context of the full numerical algorithm in which it is embedded. The relevant question is thus not whether the quantum subroutine produces an essentially exact solution at each step, but rather what level of inexactness can be tolerated without significantly degrading the convergence and final accuracy of the overall scheme. This consideration is particularly important in view of early fault-tolerant quantum computing, where the cost of tightly suppressing solver error may be substantial. As a result, the first practically useful quantum applications may well operate at a lower solver accuracy than is standard in mature classical CFD practice, provided that this does not compromise the performance of the full hybrid algorithm.

In state-of-the-art CFD applications, classical linear solvers are typically implemented in double-precision arithmetic and can be converged to an accuracy limited primarily 
by the \textit{stiffness} of the matrix and the rate of convergence of the linear solver.
Achieving comparable levels of solver accuracy with quantum computing may, however, be impractical in realistic settings. One reason for this discrepancy is the sampling error intrinsic to quantum measurement. Another contribution, which is the focus of this work, is algorithmic error, whose reduction comes at the cost of increasing circuit depth. The optimal complexity of QLSs scales as $\kappa \log \epsilon^{-1}$, where $\kappa$ is the condition number of the linearised matrix
and $\epsilon$ is the desired accuracy~\cite{Childs_etal2017,Morales_etal2025}. In CFD applications, the condition number of the discretized operators can reach values of $10^6$ or higher even for moderately sized problem instances. As a result, QLS implementations already require very deep circuits before stringent accuracy requirements are imposed, and further reductions in algorithmic error come at the cost of additional circuit depth. Therefore, while finite functional error does not preclude the use of quantum methods for CFD, the development and analysis of such applications must explicitly account for the effects of limited solver accuracy.

For QLSs based on polynomial approximations, the dominant source of functional error arises from approximating the inverse function 
$1/x$ by a polynomial over a finite spectral interval, typically  $[1/\kappa,1]$. Recent works~\cite{Childs_etal2017, Gribling_etal2024, Sunderhauf_etal2025} have studied this approximation error in detail for various polynomial constructions, explicitly characterizing how the achievable accuracy scales with the polynomial degree, which directly determines the algorithmic complexity of the corresponding quantum circuits. In particular, these recent results identify optimal polynomial approximations for a fixed polynomial degree under different error metrics. While these results provide tight bounds on the approximation error as a function of circuit complexity, the impact of such errors within hybrid quantum–classical workflows has not been systematically investigated. In the present work, we address this issue by analyzing how functional error induced by the polynomial approximation propagates through a specific hybrid CFD application.

In addition to the error arising from the polynomial approximation itself, one may also consider the effects of approximate implementations of the quantum primitives used to construct the solver. These can arise from, e.g., trimming small rotation angles, smoothing classical data for efficient state loading, or inexact block-encoding procedures. While different QLSs are typically compared in terms of their asymptotic gate complexity, the impact of these approximations has received less attention.

\subsection{Contributions}

In this work, we analyze two QLS constructions and numerically investigate the resulting trade-offs between circuit depth, width, and the performance of the hybrid quantum-classical CFD application, while also considering approximate implementations. We focus on resource estimation in the framework of fault-tolerant compilation, where the number of single-qubit rotations serves as a proxy for the computational cost. Specifically, we make the following contributions:
\begin{itemize}
    \item We numerically analyze how polynomial approximation error in a quantum linear solver propagates through a hybrid CFD workflow, showing that convergence is preserved under moderate inexactness. 
    \item We introduce an approximate version of the Cheb-LCU solver, a qubitization-based approach~\cite{Childs_etal2017} to polynomial block-encoding, and compare it to the QSVT solver~\cite{Martyn_etal2021} in circuit resources and CFD convergence.
    \item We demonstrate how the approximate Cheb-LCU variant reduces the required single-qubit rotations by over an order of magnitude, with only a moderate increase in CFD iteration count.
    \item We provide a full end-to-end implementation with all quantum code in an open repository~\cite{repo:rr, repo:classiq}.
\end{itemize}

The rest of this paper is organized as follows. In Sec.~\ref{sec:cfd} we introduce the hybrid algorithm analyzed throughout this work, and in Sec.~\ref{sec:qlss} we discuss the two quantum linear solvers. Then, in Sec.~\ref{sec:results} we present end-to-end results for a simple problem geometry, comparing the required resources of each quantum solver in view of the convergence of the full hybrid scheme. Finally, discussion and conclusions are given in Sec.~\ref{sec:discussion}. Several technical issues, such as numerical methods for Chebyshev expansions and the treatment of boundary conditions, and their implications for the performance of the hybrid solver are provided in four appendices.

\section{Hybrid classical quantum solver for 1D Nozzle flow}
\label{sec:cfd}
Flow within a converging-diverging nozzle is a common benchmark case for high-resolution shock-capturing
schemes. Here, we use the open source solver 1D-Nozzle~\cite{rollsroyce_qccfd_1d_nozzle} designed for the investigation of hybrid classical-quantum CFD workflows.
The 1-dimensional steady state equations to be solved are:
\begin{equation}
  \begin{array}{rcl}
    \rho u \frac{\partial u}{\partial x} &=& - \frac{\partial p}{\partial x}, \\[8pt]
    \frac{\partial \rho u}{\partial x} &=& 0, \\
  \end{array}
  \label{eqn-euler-1d}
\end{equation}

Whilst the solver can model transonic compressible flows, we consider low speed incompressible flows with
a constant density, $\rho=1$. The presence of the non-linear term $u\frac{\partial u}{\partial x}$ in 
\Cref{eqn-euler-1d} means that an iterative solver is needed where $u$ and $p$ are updated via:
\begin{equation}
  \begin{array}{rcl}
     u^{n+1}_i &=& u^n_i + \delta u^n_i, \\
     p^{n+1}_i &=& p^n_i + \delta p^n_i, 
  \end{array}
  \label{eqn-iter-update}
\end{equation}
where $i$ denotes the axial station within the nozzle and $n$ is the non-linear iteration counter. 
Using a uniform 1-dimensional mesh with constant axial spacing of $\delta x$, variable nozzle area $a_i$ and a fixed point iteration, 
\Cref{eqn-euler-1d} can be discretised as \cite{gupta1993development}:
\begin{equation}
  \begin{array}{rcl}
    u^n_{i-\frac{1}{2}} (\delta u^n_i - \delta u^n_{i-1}) +
    (\delta p^n_i - \delta p^n_{i-1}) &=&
    -u^n_{i-\frac{1}{2}} (u^n_i - u^n_{i-1}) -
    (p^n_i - p^n_{i-1}) = R^n_{u,i}, \\
     a_{i+1} \delta u^n_{i+1} - a_{i} \delta u^n_{i} &=&
    -(a_{i+1} u^n_{i+1} - a_{i} u^n_{i}) = R^n_{p,i}.
  \end{array}
  \label{eqn-euler-1d-disc2}
\end{equation}

\begin{figure}[h]
  \begin{center}
    \includegraphics[width=0.40\textwidth]{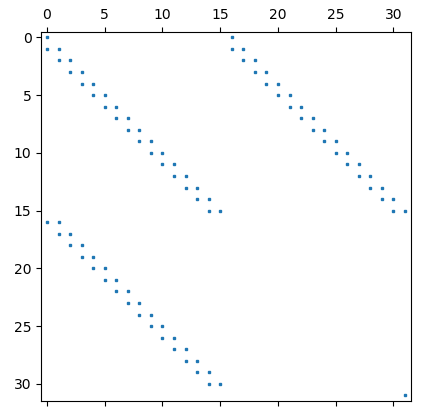}
  \end{center}
  \caption{Sparsity pattern for an incompressible coupled matrix with 16 stations.}
  \label{fig-i16-cpl-sparse}
\end{figure}

The terms $R^n_{u,i}$ and $R^n_{p,i}$ are the equation residuals, i.e. the error in the non-linear flow
field. It is easily seen that as the corrections $\delta u$ and $\delta p$ tend to zero, the non-linear 
equations tend to their solution.
\Cref{eqn-euler-1d-disc2} is a set of coupled equations that can be represented in matrix form as:
\begin{equation}
  \begin{pmatrix}
    A^n_u + B^n_{u} & D^n_x  \\
    M^n_x         & B^n_{p} \\
  \end{pmatrix}
  \begin{pmatrix}
    \vec{\delta u^n} \\
    \vec{\delta p^n}
  \end{pmatrix}
  = 
  \begin{pmatrix}
    \vec{R^n_u} \\
    \vec{R^n_p}
  \end{pmatrix}.
  \label{eqn-implicit-01}
\end{equation}
From \Cref{eqn-euler-1d-disc2}, we can see that the entries for row $i$ in each of the matrices are:
\begin{equation}
  \begin{array}{rcl}
      A^n_{u,i} &=& ( u^n_{i-\frac{1}{2}}, - u^n_{i-\frac{1}{2}}), \\
      D^n_{x,i} &=& (1, -1), \\
      M^n_{x,i} &=& (a_{i+1}, -a_i).
  \end{array}
  \label{eqn-couple}
\end{equation}
The matrix $ B^n_{u}$ contains the velocity boundary condition and its only non-zero entry is at the nozzle inlet.
Similarly, $ B^n_{p}$ is the pressure boundary condition and its only non-zero entry is at the nozzle outlet.
The sparsity pattern for a coupled matrix with 16 stations is shown in \Cref{fig-i16-cpl-sparse}.

The attraction of the coupled solver is that it gives rapid convergence. It carries a memory overhead that can
become significant for larger scale problems. For this work, it means that different QLS approximations can be
evaluated with a small number of non-linear iterations. 
In the subsequent sections we denote the linear system in \Cref{eqn-implicit-01} as $A\vec{x} = \vec{b}$,
this can hide the fact that
we are solving a correction system of the form: $A\delta\vec{x} = \delta\vec{b}$.
The scaling needed to normalize $\delta\vec{b}$ is used later to scale the QLS solution vector.
For block encoding, $A$ may also need to be scaled to have a max norm $\le 1$, the corresponding scaling
factor can be handled similarly.

\section{Quantum Linear Solvers}
\label{sec:qlss}

Following the previous section, the quantum part of the hybrid scheme solves a linear system of equations, defined by a matrix $A$ and a right-hand-side vector $\vec{b}$:
\begin{equation}
    A\vec{x} = \vec{b},
\end{equation}
where $\vec{x}$ is the solution. For convenience, we assume that the matrix size is $2^n\times 2^n$.
The two QLSs we consider are based on the idea of block-encoding--- we assume that the matrix $A$ is embedded in a larger unitary matrix, 
of size $2^{n+m}\times 2^{n+m}$: 
\begin{equation}
    U_{A} = \begin{pmatrix}
        A/s & * \\
        * & *
    \end{pmatrix},    
\end{equation}
where $m$ is the number of embedding qubits (referred to hereafter as "block" qubits), and $s$ is some scaling factor. 
To ensure unitarity of the matrix, we must have $s \ge \|A\|$, where $\|\cdot\|$ denotes the max norm. The value of $s$ is strongly dependent on the block-encoding method~\cite{lapworth2024}.
The singular values of the block encoded matrix lie in the range $[1/\kappa_s,1]$, with an effective condition number
\begin{equation}
  \kappa_s = \frac{s}{\lambda_{\min}},
  \label{eqn:be-subnorm}
\end{equation}
where $\lambda_{\min}$ is the smallest singular value of $A$.

We consider quantum matrix inversion routines that are based on polynomial singular value transformation~\cite{martyn2021grand}: 
Given $U_A$ as an input, and an odd polynomial approximation
\begin{equation}
  P(y) \approx \frac{1}{y}, \qquad y\in \left[-1,-\frac{1}{\kappa_s}\right] \lor \left[\frac{1}{\kappa_s},1\right], 
  \label{eqn-qsvt-subnorm05a}
\end{equation}
we construct an approximated block-encoding\footnote{The notation of the matrix polynomial should be
understood here in the sense of singular value transformation. Given the Singular Value Decomposition of $A$,
$A=W\Sigma V^{\dagger}$, where $\Sigma$ is a diagonal matrix containing the singular values of $A$ and $W$, $V$ are unitary matrices,
the odd polynomial singular value transform of $A$ is defined as  $P(A) \equiv W P(\Sigma) V^{\dagger}$,
with $P(y)$ being an odd polynomial. In particular, for the inverse function, using the fact that $A^T=V\Sigma W^{\dagger}$ 
we get $P(A^T) \approx V\Sigma^{-1} W^{\dagger} = A^{-1}$.} of the inverse
\begin{equation}
   U_{P\left(A^{T}/s\right)} = \begin{pmatrix}
        cP\left(A^T/s\right) & * \\
        * & *
    \end{pmatrix} \approx 
    \begin{pmatrix}
        \left(cs\right)A^{-1} & * \\
        * & *
    \end{pmatrix}=U_{A^{-1}},
\label{eq:be_a_inv}
\end{equation}
where $c$ is a scaling factor that depends on the specific details of the QLS.
In turn, this unitary can be applied on a prepared state of the right-hand-side vector to get a solution embedded in a quantum state:
\begin{equation}
    U_{P\left(A^{T}/s\right)} \cdot |b\rangle_n|0\rangle_{m_{\rm inv}} \approx
    \left(\left(cs\right)\|A^{-1} |b\rangle\|\right)|x\rangle_n|0\rangle_{m_{\rm inv}} + |\widetilde{\perp}\rangle,
    \label{eq:be_state}
\end{equation}
where $m_{\rm inv}>m$ is the size of block variable, $|\widetilde{\perp}\rangle$ is some  unnormalized state that 
is orthogonal to the zero block state (the $\widetilde{}{\,}\,$ indicates an unnormalized vector, whereas all the other $|\cdot\rangle$ states are normalized).
Note that the size of the inverse block variable $m_{\rm inv}$  depends on the specific details of the QLS.
The values of $c$ and $m_{\rm inv}$ for the two solvers used in this paper are discussed in detail in Sec.~\ref{sec:resources}.
Note that the scaling factor $cs$ introduces a multiplicative factor for the solver's complexity,
as it is directly related to the success probability of measuring the solution,
$\left(cs\right)^2 \|A^{-1} |b\rangle\|^{2}$,
or equivalently to the cost of amplitude amplification.

We can write the approximated inverse function as an expansion of odd Chebyshev polynomials with alternating signs\footnote{The alternating signs for the 
odd coefficients were confirmed numerically, for all the polynomial expansions we use. A motivation for why this is an expected behavior is given in Appendix~\ref{app:approx_cheb_approx_coeffs}}:
\begin{equation}
  \frac{1}{y} \approx P(y)=  \sum^{(d-1)/2}_{j=0} (-1)^j a_j T_{2j+1}(y),  \quad a_j\geq 0 , \qquad y\in [-1,-1/\kappa] \lor [1/\kappa,1], 
  \label{eq:f_inv}
\end{equation}
with $T_k$ being the $k$-th Chebyshev polynomial. In Sec.~\ref{sec:implementation_details} we discuss the specific Chebyshev expansion we take for the CFD application, whereas in Appendix \ref{app-large-eval} we explore the implications of different Chebyshev expansions. For the rest of the paper, we assume that the polynomial degree follows $d=2(2^l-1)+1=2^{l+1}-1$, for some integer $l$, since it allows an easy one-to-one comparison between the two solvers.

The two QLSs we use are (1) the well-known Quantum Singular Value Transform (QSVT) method,
% ~\footnote{Both methods apply a Quantum Singular Value Transform on a matrix, however, 
% we keep the usual naming in the literature, using QSVT to describe the first method that is based on Quantum Signal Processing.}
and (2) a more direct approach that applies an explicit block-encoding of the Chebyshev polynomial, using Linear Combination of Unitaries (LCU), designated hereafter as Cheb-LCU. In addition, we consider an approximated version of the Cheb-LCU approach.

\subsection{QSVT solver}
\begin{figure}[h!]
    \centering
    \includegraphics[width=0.95\linewidth]{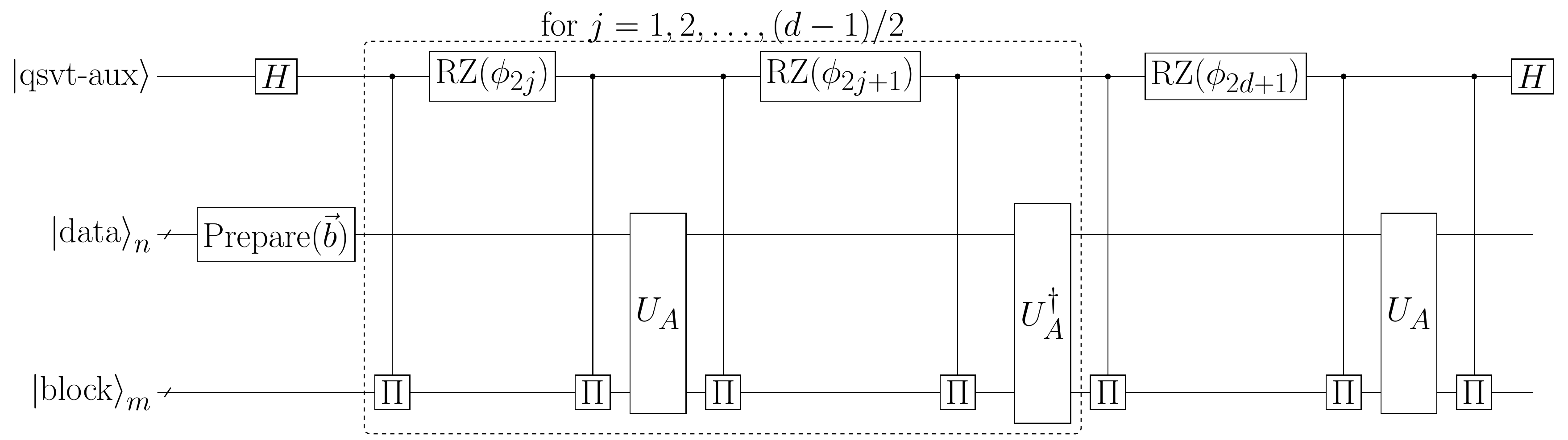}
    \caption{A QSVT circuit for applying an odd polynomial transformation, with degree $d$, on a block-encoded matrix.}
    \label{fig:qsvt}
\end{figure}

The QSVT solver is based on Quantum Signal Processing~\cite{Martyn_etal2021}. In this approach, the desired polynomial transformation of the singular values is applied directly to a block-encoding of the system matrix (we assume hereafter that $A$ is a real valued matrix, therefore $U^{\dagger}_A=U_{A^T}$).
More specifically, given block-encodings of $A$ and $A^T$, one constructs a block-encoding of $P(A^T/s)$ using a single additional block qubit together with a sequence of single-qubit $R_Z$ rotations, parameterized by the so-called QSVT phase angles. These rotations are interleaved with applications of $U_{A}$, $U^\dagger_{A}$, and controlled-$\Pi$ operations, where $\Pi$ denotes the reflection about the block state $|0\rangle_m$,
\begin{equation}
    \Pi \equiv \mathcal{I}_m-2|0\rangle_m{}_m\langle 0|,
    \label{eq:pi}
\end{equation}
with $\mathcal{I}_m$ being the identity matrix on $m$ qubits.
The full unitary operation for an odd polynomial of degree $d$ reads~\cite{Martyn_etal2021}:
\begin{equation}
  \Pi_{\phi_0} U_A \left[ \prod\limits_{k=1}^{(d-1)/2} \Pi_{\phi_{2k-1}} U_{A}^{\dagger} \Pi_{\phi_{2k}} U_{A}\right]  
  =
   \begin{pmatrix}
    P(A^T/s) & * \\
    *           & *
  \end{pmatrix}
  \label{eqn-martyn32}
\end{equation}
where $\Pi_{\phi}$ is the projector controlled phase shift operator~\cite{Martyn_etal2021};
see also a schematic layout of the corresponding quantum circuit in Fig.~\ref{fig:qsvt}.
The resulting model realizes an odd polynomial transformation of the singular values of the encoded matrix, and therefore can be used to approximate the inverse map required for the linear-system solver.
Following~Eq.\eqref{eqn-martyn32} and Fig.~\ref{fig:qsvt}, we can see that the practical cost of this construction is governed by both the polynomial degree and the complexity of the underlying block-encoding calls. A more detailed discussion of the resource requirements of the QSVT approach is given in subsection~\ref{sec:resources}.

\subsection{Cheb-LCU solver}
The second solver avoids the need to compute QSP angles, and relies on the so-called Qubitization approach (based on the Szegedy walk operator~\cite{Szegedy2004}).
Using $U_{A}$ and $\Pi$, we can define a set of operators that block-encodes the odd {\it Chebyshev} polynomial transformations of $A^T/s$ (Section 8.2 in Ref.\cite{Linlin_notes}):
\begin{equation}
U_{T_{2j+1}(A^T/s)} = \begin{pmatrix}
        T_{2j+1}(A^T/s) & * \\
        * & *
    \end{pmatrix}=
    U^{\dagger}_{A}\Pi \left(U_{A} \Pi U^{\dagger}_{A} \Pi \right)^j
    \equiv 
    W^{(2j+1)}_{U_{A^T}}
    .
\end{equation}
Therefore, we can construct the desired block-encoding of the sum of Chebyshev polynomials in Eq.~\eqref{eq:f_inv},
using an extra $l$ block qubits, via the simple  PREPARE-SELECT routine:
\begin{equation}
     U_{P(A^T/s)} = \text{PREPARE}^{\dagger}\cdot  \text{SELECT}\cdot\text{PREPARE},
\end{equation}
with 
\begin{equation}
  \text{PREPARE} \equiv \left(\sum^{2^l-1}_{j=0} \sqrt{a_j/||\vec{a}||_1} |j\rangle\langle 0| + \sum_{j=0,k=1}^{2^l-1} g_{jk} |j\rangle\langle k| \right) \otimes \mathcal{I}_{n+m},
  \label{eq:prepare}
\end{equation}
\begin{equation}
    \text{SELECT} \equiv  \sum^{2^l-1}_{j=0} |j\rangle\langle j| \otimes \left(-1\right)^{j}W^{(2j+1)}_{U_{A^T}} =
\left(\mathcal{I}_l \otimes U_{A^T}\Pi\right)  \sum^{2^l-1}_{j=0} |j\rangle\langle j| \otimes \left(-W^{(2)}\right)^j ,
\label{eq:select}
\end{equation}
with $||\vec{a}||_1 = \sum_j |a_j|=\sum_j a_j$ and $W^{(2)} \equiv \left(U_{A}\Pi U_{A}^{\dagger} \Pi\right)$. 

\begin{figure}[h!]
    \centering
        \includegraphics[width=0.95\linewidth]{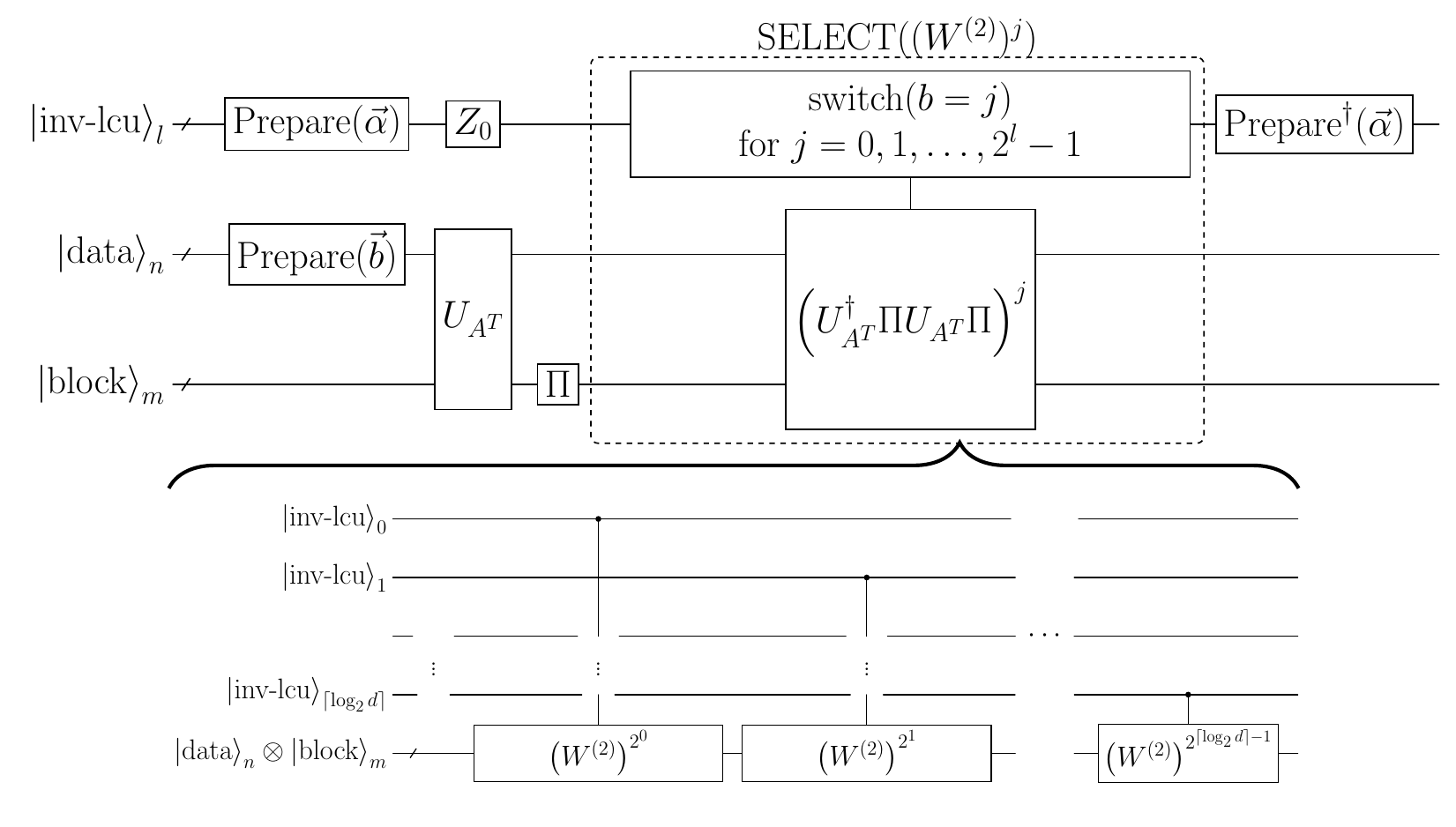}
    \caption{The layout of the Chebyshev LCU model for matrix inversion. The $Z$ gate on the first inversion block qubits
    implements the alternating signs in Eq.~\eqref{eq:f_inv}.
    The lower part expands the implementation for the select powers of $W^{(2)} = \left(U_{A}\Pi U^{\dagger}_{A} \Pi\right)$ function. Note that in practice
    when controlling this function, we can remove the controls over $U_{A}$ and $U^{\dagger}_{A}$.}
    \label{fig:cheb_lcu}
\end{figure}

Since the \text{SELECT} operation is a series of controlled-powers of the same unitary $W^{(2)}$, we can implement it using single controlled operations of that unitary; see Fig.~\ref{fig:cheb_lcu} for the quantum model layout. Furthermore, in all these controlled operations we can "skip" controlling over $U_{A}$ and $U^{\dagger}_{A}$ since their product is the identity. This leaves us with a sequence of operations similar to the QSVT, 
containing controlled-$\Pi$, $U_{A}$ and $U_{A}^{\dagger}$. In the next section we discuss in detail this similarity, and where the two QLSs differ.

\subsection{Theoretical resource analysis}
\label{sec:resources}

Let us now summarize the different properties of each solver. We recall the assumption that $d=2^{l+1}-1$, which allows an easy comparison between the methods. Both methods are constructed with the same number of calls for $U_{A}$ and $\text{control}-\Pi$ (where we count calling $U_{A}^{\dagger}$ as calling $U_{A}$). The QSVT includes $d$ $R_Z$ rotations on an extra block qubit, whereas the Cheb-LCU includes two state preparations for the coefficients $a_j$ on extra $l$ block qubits. Without going into implementation details, preparing a state
of $2^l$ arbitrary amplitudes is expected to include $O(2^l)$ rotations; However, as we show in Sec.~\ref{sec:results}, in practice, the resources for the state preparation can be
reduced significantly--- this is one of the main results of this paper. 

Finally, concerning the scaling factor $c$ that appears in Eq.~\eqref{eq:be_a_inv}, in QSVT this value originates in the fact that we can only implement polynomial bounded in absolute value by 1, thus this 
solver encodes the polynomial transformation with $c^{\rm QSVT}=\|P\|_{\max}\equiv\max_{y\in[-1,1]}P(y)$.
In the  Cheb-LCU solver, we get a normalization from the PREPARE routine of the LCU, which results in $c^{\rm Cheb-LCU}=\|\vec{\alpha}\|_{1}$, see Eq.~\eqref{eq:prepare}. 
In general, we have $c^{\rm QSVT}\leq c^{\rm Cheb-LCU}$, however in Ref.~\cite{Gribling_etal2024} it was shown, that for some polynomial expansions, in particular the one we use in this work (see Theorem 17 therein), $c$ for Cheb-LCU (and thus for QSVT) is optimal up to a factor of $\log(\kappa_s^2/\epsilon)$, with $\epsilon$ being the polynomial approximation error. In Table.~\ref{tab:QLSs} we summarize the properties of the two QLSs.

In this work we consider the property which makes the main difference between the solvers, the $d$ calls for RZ rotations in QSVT vs. preparing the
$\vec{a}$ coefficients of the Chebyshev polynomial on a quantum variable in Cheb-LCU. 
In particular, we focus on how the algorithmic approximation of the latter can reduce the total number of rotations and its effect on the full hybrid scheme performance. This is presented in the next Section.
Finally, we note that another apparent difference between the methods is the classical pre-computation of the QSVT angles; however, recent work~\cite{nlft_2025} has shown that this step is not a crucial bottleneck for QSVT algorithms.

\begin{table}[t]
\centering
\begin{tabular}{lccccc}
\toprule
QLS & $U_{A}$ & ctrl-$\Pi$ & {\bf other} & $m_{\rm inv}$ & $c$\\
\midrule
QSVT & $d$ & $d$  & $d$  {\bf RZ gates with the QSVT angles} & 1& $\|P\|_{\max}$\\
Cheb-LCU & $d$ &$d$  & {\bf 2 State preparations of $d=2^l$ for $\vec{a}$}  & $l+1$& $\|\vec{a}\|_{1}$\\
\bottomrule
\end{tabular}
\caption{A summary table for the properties of QLSs used in the CFD application. The bold entries indicate
the main difference between the methods, and it is in the focus of our study in terms of resource estimation.}
\label{tab:QLSs}
\end{table}

\subsection{Approximated Cheb-LCU}

The Cheb-LCU approach offers a simple way to reduce depth at the expense of the QLS accuracy. This can be done
by approximating the PREPARE block of the Chebyshev coefficients at the beginning and at the end of the model. Comparing to the QSVT solver, these blocks
exchange the RZ rotations, as can be seen from Table.~\ref{tab:QLSs}. If the coefficients vector $\vec{a}$ forms a continuous function, it is plausible that there is an 
efficient implementation for loading it into a quantum variable. Even if such an implementation is not available, one can consider an approximate implementation by binning the domain (i.e., using a piecewise-constant / step-function approximation), possibly after smoothing. Then, one needs to analyze how such approximation affects the QLS solution. In the next section we explore
this issue experimentally, showing that the number of resources for the Chebyshev coefficients loading (PREPARE) can be reduced significantly, with very little trade off for the CFD accuracy.
The specific details of approximated data loading is given in Sec.~\ref{sec:implementation_details}.

\section{End-to-end runs for 1D nozzle geometry}
\label{sec:results}

In this section we present results for the full nonlinear CFD application. We consider the performance of the two QLSs, exploring how algorithmic approximation 
affects the overall convergence of the CFD routine and the quantum resources. Our main focus is approximations of the Cheb-LCU solver, as it 
is less studied compared to QSVT. First, we discuss several important {\it implementation details} for the QLSs in Sec.~\ref{sec:implementation_details}, and then,  in Sec.~\ref{sec:comparison}, present results of the CFD run along with resource comparison.

\subsection{Implementation details}
\label{sec:implementation_details}

We briefly explain the technical details of the three important elements of the implementation: block-encoding implementation, polynomial expansion for the inverse function, and approximated loading of the Chebyshev coefficients; More details are provided in Appendices~\ref{app:be}, \ref{app:chen_approx}, and \ref{app:sp}, respectively.

{\it Block-encoding $A$}:  From Fig.~\ref{fig-i16-cpl-sparse}, we can see that the matrix $A$ has a finite number of non-vanishing diagonals. These diagonals contain arbitrary entries.
In this work we implement the block-encoding of $A$ according to Ref.~\cite{Lapworth&Sunderhauf2025}, where we perform state-loading of each diagonal, and then use controlled adders for "shifting" them to the desired diagonal index. If $A$ contains $K$ non-vanishing diagonals, we designate for $i=0,\dots K-1$:
\begin{itemize}
    \item $h^{(i)}$ - the position of the $i$-th diagonal, an integer index where positive/negative values refer to diagonals below/above the main one.
    \item $\vec{z}^{(i)}$ - the values of the $i$-th diagonal, a vector of real numbers of size $2^n-|h^{(i)}|$. 
    \item $w^{(i)}\equiv\|\vec{z}^{(i)}\|_{\max}$ the max norm of the $i$-th diagonal.
\end{itemize}
Then, the encoding procedrue follows a PREPARE-SELECT routine, where the PREPARE corresponds to loading the diagonals norms
$\left(\sqrt{w^{(0)}},\sqrt{w^{(1)}},\dots,\sqrt{w^{(K-1)}}\right)/\sqrt{\sum^{K-1}_{i=0} w^{(i)}}$ on $k\equiv \lceil \log_2K\rceil$ qubits, 
and the SELECT operation selects unitaries that block encodes matrices with a single diagonal. See more details in Appendix~\ref{app:be} and Fig.~\ref{fig:be}
for the model layout. This method has $k+1$ block qubits, and
a scaling factor of $s=\sum^{K-1}_{i=0}w^{(i)}=\sum^{K-1}_{i=0}\|\vec{z}^{(i)}\|_{\max}$.

{\it Chebyshev polynomial expansion of $1/y$}: 
For the polynomial approximation of $1/y$ in Eq.~\eqref{eq:f_inv} we follow Ref.~\cite{Gribling_etal2024}, which provides an odd polynomial, which minimizes the relative error
$\max_{y\in [-1,-1/\kappa] \lor [1/\kappa,1]}|yP(y)-1|$ for a given degree $d$. We find that other polynomial expansions, for example one that minimizes the uniform error 
$|P(y)-1/y|$~\cite{Sunderhauf_etal2025}, are irrelevant for our CFD application; see detailed discussion in Appendix~\ref{app:chen_approx}.

\begin{figure}[h!]
\centering

% Row 1
\begin{minipage}[t]{0.49\linewidth}
  \centering
  \includegraphics[width=\linewidth]{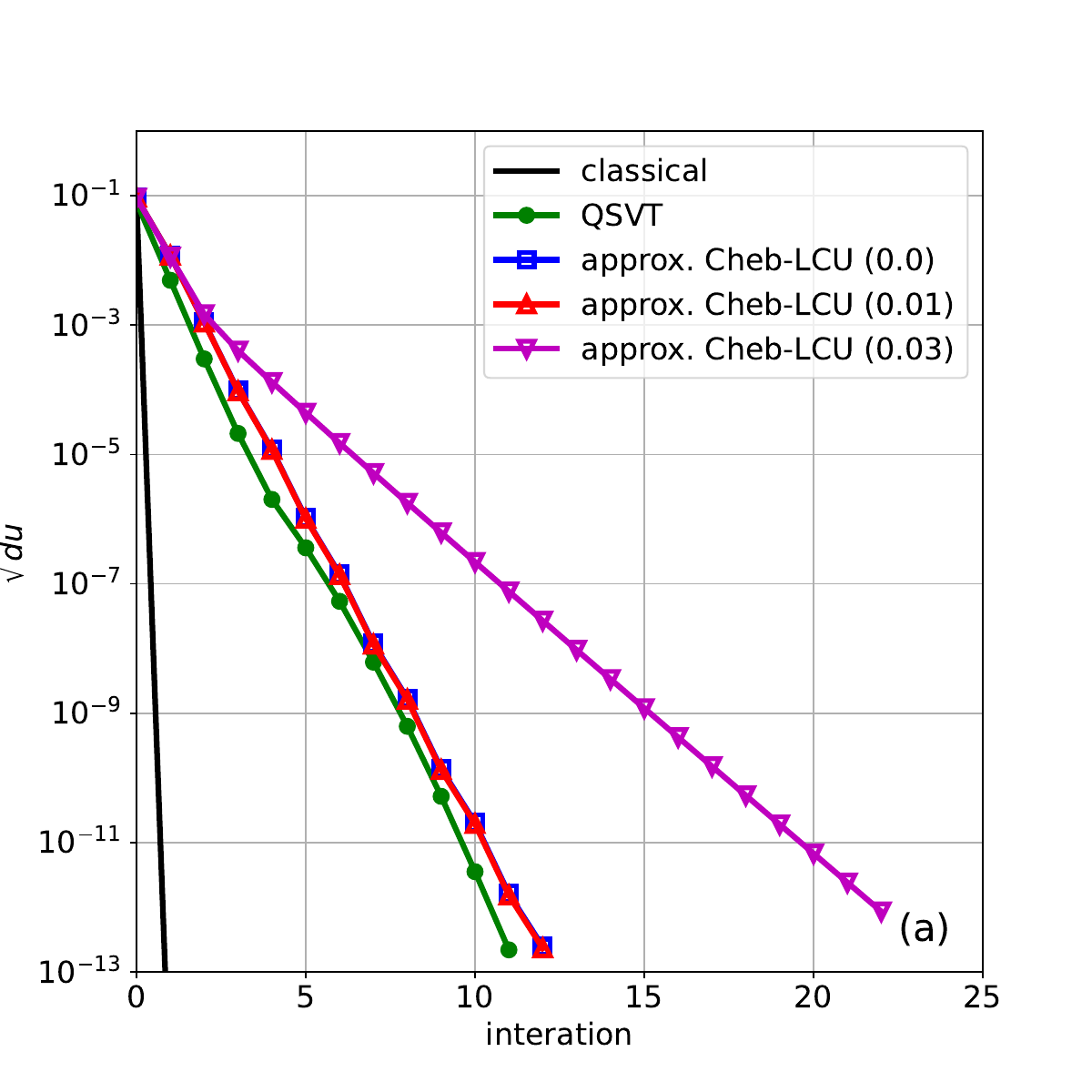}
\end{minipage}\hfill
\begin{minipage}[t]{0.49\linewidth}
  \centering
  \includegraphics[width=\linewidth]{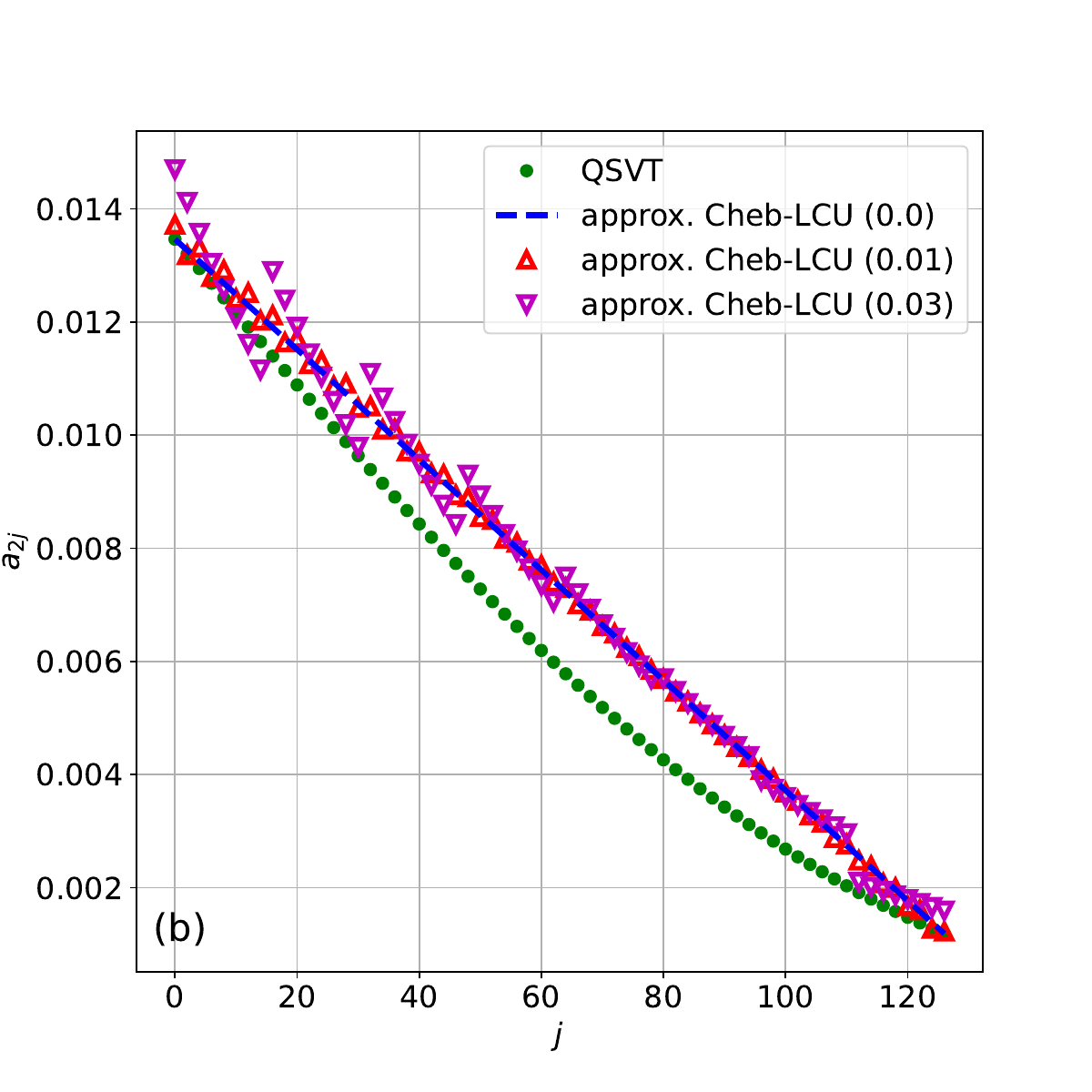}
\end{minipage}

\vspace{0.6em}

\newcommand{\rowh}{1.2cm} 

% Row 2
\begin{minipage}[c]{0.49\linewidth}
  \centering
  \includegraphics[width=\linewidth]{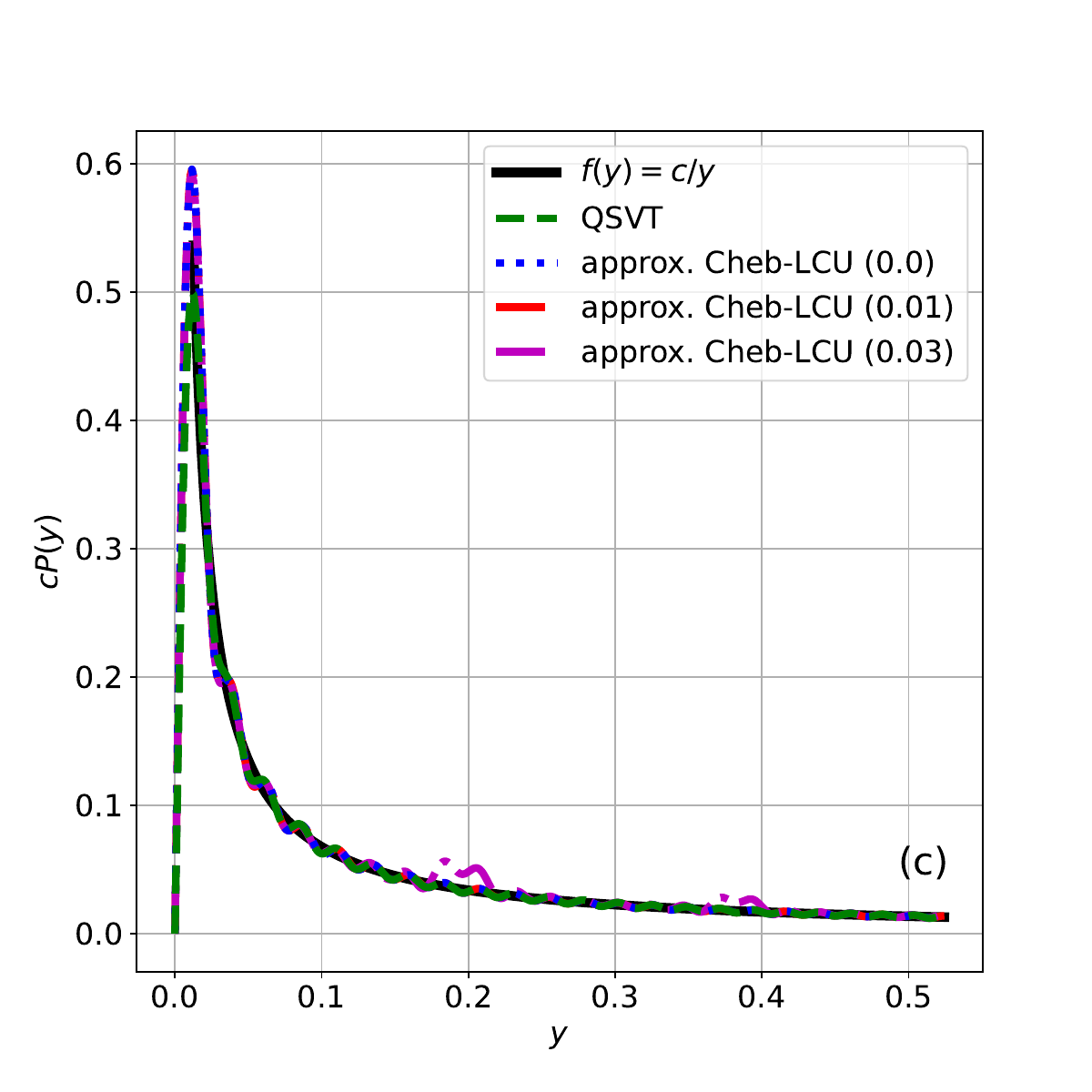}
\end{minipage}\hfill
\begin{minipage}[c]{0.49\linewidth}
  \centering
  \small
  \begin{tabular}{lcc}
    \hline
    Method & \# angles & $c$ \\
    \hline
    QSVT & 256 & $1.35e^{-2}$ \\
    approx. Cheb-LCU (0.0) & 254 & $1.26e^{-2}$ \\
    approx. Cheb-LCU (0.01) & 66 & $1.26e^{-2}$\\
    approx. Cheb-LCU (0.03) & 12 & $1.26e^{-2}$ \\
    \hline
  \end{tabular}
  \captionof{table}{Comparison of number of rotations for each solver. For QSVT this refers only to count
  of QSVT angles, whereas for the Cheb-LCU only to the angles required for the PREPARE$(a)$ and its inverse. The prefactor $c$,
  which is related to the success probability (see Eq.~\eqref{eq:be_state}), is included as well for each solver.}
  \label{tab:resources}
\end{minipage}

\caption{Comparison on the performance of the QSVT solver and approximated Cheb-LCU solvers. (a) The convergence histories, 
for the velocity correction, of each solver, including the reference history using a classical (scipy) linear solver. The other three panels show the 
characteristics of the solvers at iteration 9. (b) The Chebyshev polynomial's positive coefficients $a_{2k}$ used for each solver
(the negative ones follow the same trend). (c) Resulting Chebyshev polynomial approximation to the target function $c/y$ over the interval $[\lambda_{\min}, \lambda_{\max}]$.
All curves use the same scaling, $c=0.5c^{\rm QSVT}=0.5\|P\|_{\max}^{-1}$, enabling a one-to-one comparison. (d) The resource comparison for the different quantum solvers, including the actual values of the polynomial block-encoding pre-factor $c$.}
\label{fig:qsvt_vs_cheb}
\end{figure}

{\it Approximated loading of the Chebyshev coefficients:}
Examining the Chebyshev polynomial coefficients $\left\{a_j\right\}$ in our examples, we find that indeed the coefficients follow some smooth curve, see blue dots in the upper-right panel of Fig.~\ref{fig:qsvt_vs_cheb} (for clearer visualization, we plot only the coefficients that enter with a positive sign into the expansion in Eq.~\eqref{eq:f_inv}, with index $j=2k$.).
There are various algorithms for loading smooth arbitrary functions, e.g., based on QSVT~\cite{Gilyen_etal2019} or Generalized-QSP~\cite{Motlagh&Wiebe2024}. 
However, in this work we take a more naive approach. First, we fit the positive and negative coefficients with a linear function, simply 
drawing a straight line between the first and last point, see dashed blue line in the upper-left panel of Fig.~\ref{fig:qsvt_vs_cheb}. This makes sure that
the approximated coefficients do not become negative. While such fitting can be then used to load linear amplitudes efficiently, we take a different approach, 
and use an approximate state-loading method for arbitrary data, according to Ref.~\cite{Marin-Sanchez_etal2023}. For more details see  Appendix~\ref{app:sp}. 

Next, we present how the approximated Cheb-LCU solver effects the hybrid CFD solution, in terms of convergence and resources.

\subsection{Explicit Resource Estimation}
\label{sec:comparison}

We treat a small example of a 1D problem of gird size 8, in which the correction iteration with a linear solver 
maps to a matrix $A$ of dimensions $2^4\times2^4$, see Fig.~\ref{fig-i16-cpl-sparse}. We run the full hybrid-CFD solver as described in Sec.~\ref{sec:cfd}. The classical part of the solver is available in
a public repository~\cite{repo:rr}, and the quantum part, developed and implemented using Classiq's Qmod language and Synthesis engine, is available in Classiq's open library~\cite{repo:classiq}.
The block-encoding of $A/s$ results in an effective condition number of $\kappa_s\approx 80$, for all iterations. Following Ref.~\cite{Gribling_etal2024}, if we would like to
use the polynomial approximation in Eq.~\eqref{eq:f_inv} to get an error of $|yP(y)-1|< 1e{-6}$, for example, then we should choose a polynomial degree of 1785.
However, for our runs, we choose a smaller degree, fixing  $d=255=2\cdot2^{7}-1$, which corresponds to an error bound of $|yP(y)-1|< 0.071$ for the function approximation, see derivation in Appendix~\ref{app:chen_approx}.

In Figure~\ref{fig:qsvt_vs_cheb}(a) we present the convergence history of the CFD algorithm for different types of linear solvers for the correction step. 
For reference, we include a purely classical scenario, using a classical scipy matrix inversion, which converges after 3 iterations
(this is a specific feature of this test-case that would require the QLS to deliver double precision accuracy to match). 
The QSVT solver, 
or an exact Cheb-LCU solver (not shown), gives convergence after 11 iterations. The green dots in Fig.~\ref{fig:qsvt_vs_cheb}(b) show an example (taken from the 9th iteration) of the decay of the Chebyshev coefficients with increasing index, and the green dashed line in Fig.~\ref{fig:qsvt_vs_cheb}(c) presents the resulting polynomial with respect to the target function $\sim 1/y$. As we can see, up to some small fluctuation, the polynomial approximation follows the target function, but overshoots at the smallest eigenvalues, near $1/\kappa_s$. The reason why this discrepancy does not affect the result is clear, as the overall convergence is not sensitive to the small part of the spectrum; See detailed discussion in Appendix~\ref{app-large-eval}.

Next, we analyze the approximated Cheb-LCU performance. As a first step, we just apply a linear fit to the Chebyshev coefficients (blue dashed line in Fig.~\ref{fig:qsvt_vs_cheb}(b)), and load the interpolated coefficients with an exact state preparation. This approach is designated by the label approx. Cheb-LCU (0.0) in all
 figure panels. As can be seen, the simple linear approximation of the curve barely affects the polynomial approximation (blue dotted line in the Fig.~\ref{fig:qsvt_vs_cheb}(c)) or the algorithmic convergence (blue line with square markers in Fig.~\ref{fig:qsvt_vs_cheb}(a)). However, since we take a naive state preparation this approach does not give any resources reduction, see Table~\ref{tab:resources}, it just demonstrates how smoothing and small modifications of the Chebyshev coefficients barely affect the results.

Finally, we consider non-exact state preparation of the Chebyshev coefficients. We take the coefficients obtained from the linear fit and apply the approximate state-loading procedure described in Sec.~\ref{sec:implementation_details}. We consider two examples with different error bounds for the approximate coefficients, $\vec{a}^{\rm approx}$, satisfying $\|\vec{a}-\vec{a}^{\rm approx}\|_2<\epsilon^{\rm sp}$, with $\epsilon^{\rm sp}=0.01$ and $0.03$. As shown in Fig.~\ref{fig:qsvt_vs_cheb}, for an error bound of 0.01 the CFD result changes only slightly, while the resources required for state preparation are reduced by a factor of 4 (red lines and triangle up markers). Even more interestingly, for an error bound of 0.03, which requires very few rotations to load the Chebyshev coefficients, the CFD algorithm still converges, although it requires about twice as many iterations (magenta lines and triangle down markers).
This feature is discussed in \Cref{app-large-eval}. As a final remark, the resources reported in Table~\ref{tab:resources}, given in terms of the number of angles, should be interpreted together with the pre-factor $c$ of each method. However, since we find $c^{\rm QSVT}\approx c^{\rm Cheb-LCU}$, this does not significantly affect the comparison.

As long as the linear solver resolves the large-eigenvalue regime with sufficient accuracy, the full CFD application still converges; moreover, the more accurate the inversion is in that regime, the faster the overall convergence. In particular, we see that even rather large perturbations of the expansion coefficients (e.g., with $\epsilon^{\rm sp}=0.03$), are translated only into moderate distortions of the resulting polynomial, specifically in the part of the spectrum to which the CFD iteration is comparatively insensitive. This explains why the full algorithm remains convergent despite a substantial reduction in the precision of the state preparation. At the same time, the slower convergence observed for larger $\epsilon^{\rm sp}$ shows that inaccuracies in the coefficient loading are not irrelevant: they are tolerated to the extent that they do not significantly degrade the approximation. In Appendix~\ref{app:approx_cheb_approx_coeffs} we discuss in more detail the relation between the Chebyshev coefficient, and the resulting polynomial.

\section{Discussion}
\label{sec:discussion}

In this work, we have examined the end-to-end performance of a hybrid classical-quantum algorithm for CFD.
Realizing a practical advantage in such frameworks requires progress in several directions, including efficient feedback between classical and quantum processing, scalable encoding of large classical datasets, and quantum linear solvers that can be implemented with manageable resource costs. It is therefore important to disentangle the different ingredients of the workflow and to assess the effect of the approximations made at each stage separately, while still evaluating their impact within the full hybrid end-to-end scheme. Here, we have focused on approximations in the quantum linear solver itself, asking to what extent its circuit depth can be reduced while still preserving convergence of the CFD algorithm.

In particular, we have focused on polynomial-transformation-based solvers and found that low-degree polynomials can be sufficient, in exchange for a manageable increase in the number of CFD iterations. Moreover, we have introduced an approximate solver based on a direct Chebyshev polynomial block-encoding, which further reduces resource requirements while maintaining convergence of the overall CFD algorithm. We have also shown that this approach requires fewer resources than QSVT-based solvers, which currently represent the standard approach for polynomial quantum linear solvers. More broadly, the finding that hybrid CFD workflows can tolerate moderate QLS inaccuracy is encouraging for the feasibility of embedding quantum linear solvers in practical scientific computing applications. It suggests that the stringent solver accuracy often assumed in worst-case complexity analyses may be relaxed in practice, potentially reducing the fault-tolerant overhead required for a practical computational benefit.

The resource reduction we identify refers specifically to the way of implementing the polynomial transformation, namely to the comparison between direct loading of the Chebyshev coefficients into a quantum variable and a QSP-based construction, in which the corresponding QSVT rotation angles must be computed and implemented. In a complete QLS, however, additional single-qubit rotations may also arise, for example from the block-encoding itself, and these contributions can dominate the overall cost. Consequently, our results are especially compelling in settings where the block-encoding can be realized with relatively small fault-tolerant overhead.

While in this work we have focused on a direct route to approximate polynomial transformations via the Cheb-LCU approach, our findings suggest that related approximation strategies within QSVT may also be worth exploring. In particular, it would be interesting to investigate whether modifying or rounding some of the QSVT angles can reduce resources while still providing a polynomial approximation sufficient for acceptable solver performance. Another promising direction would be to extend the Cheb-LCU approach to general polynomial degrees, beyond the form $2^{l+1}-1$ considered here, thus bringing it closer to QSVT-based QLSs, which naturally accommodate arbitrary polynomial degree.

In this work, we have focused on a small-scale test case. The effective condition number considered here, $\kappa_s \approx 80$, is substantially lower than those encountered in larger-scale CFD problems, which can reach $10^6$ or beyond; understanding how the observed trade-offs scale with condition number therefore remains an important direction for future work. Nevertheless, we expect the main qualitative conclusions to extend to larger and more general problems: (1) the full CFD scheme can continue to converge under a polynomial transformation with moderate relative error, and (2) an approximate Cheb-LCU construction can require fewer resources than QSVT for realizing such an approximation. As discussed in Appendix~\ref{app-large-eval}, the essential reason is that the polynomial approximation provides a good estimate of the inverse for the largest eigenvalues of the linear system. This mechanism should be relevant for other discretized problems as well.

\begin{acknowledgments}
We thank O. Samimi-Golan, E. Cornfeld, L. Preminger, and G. Kishony for helpful discussions.
\end{acknowledgments}

\newpage
\appendix
\input{include/appendix-cfd}
\input{include/appendix-qls}

\bibliographystyle{unsrt}
\bibliography{cfd_qls_bib}% 

\end{document}

%% file: include/appendix-cfd.tex
%
% CFD features - boundary conditions and 1/x fit
% ----------------------------------------------
\section{Practical CFD considerations}
\label{app-practical}

In this section, we highlight two practical considerations for quantum linear equation
solvers that can often be overlooked.

% 1/x at RHS
% ----------
\subsection{Approximation of $1/y$ for large eigenvalues}
\label{app-large-eval}

The focus of many approximations for $1/y$ is the accuracy of the fit
near $y=1/\kappa_s$, and many schemes apply a uniform error bound over the
region $[1/\kappa_s, 1]$.
Within CFD solvers, the impact of errors in the $1/y$ approximation on the
outer non-linear convergence can depend on locally high errors away from
$1/\kappa_s$.

At the end of each linear solution, the resulting state vector can be expressed in terms
of the eigenvectors of $A$ (for non-Hermitian matrices this can be either the left or right eigenvectors):

\begin{equation}
    \ket{x} = \sum_{i=0}^{N-1} \alpha_i \ket{\psi_i},
    \label{eqn:app1}
\end{equation}
with $N=2^n$ and the order of $\ket{\psi_i}$ based on ascending absolute values of the eigenvalues.

\begin{figure}[htb]
  \centering
  \captionsetup{justification=centering}
    \begin{subfigure}[b]{0.40\textwidth}
      \centering
      \includegraphics[width=.9\textwidth]{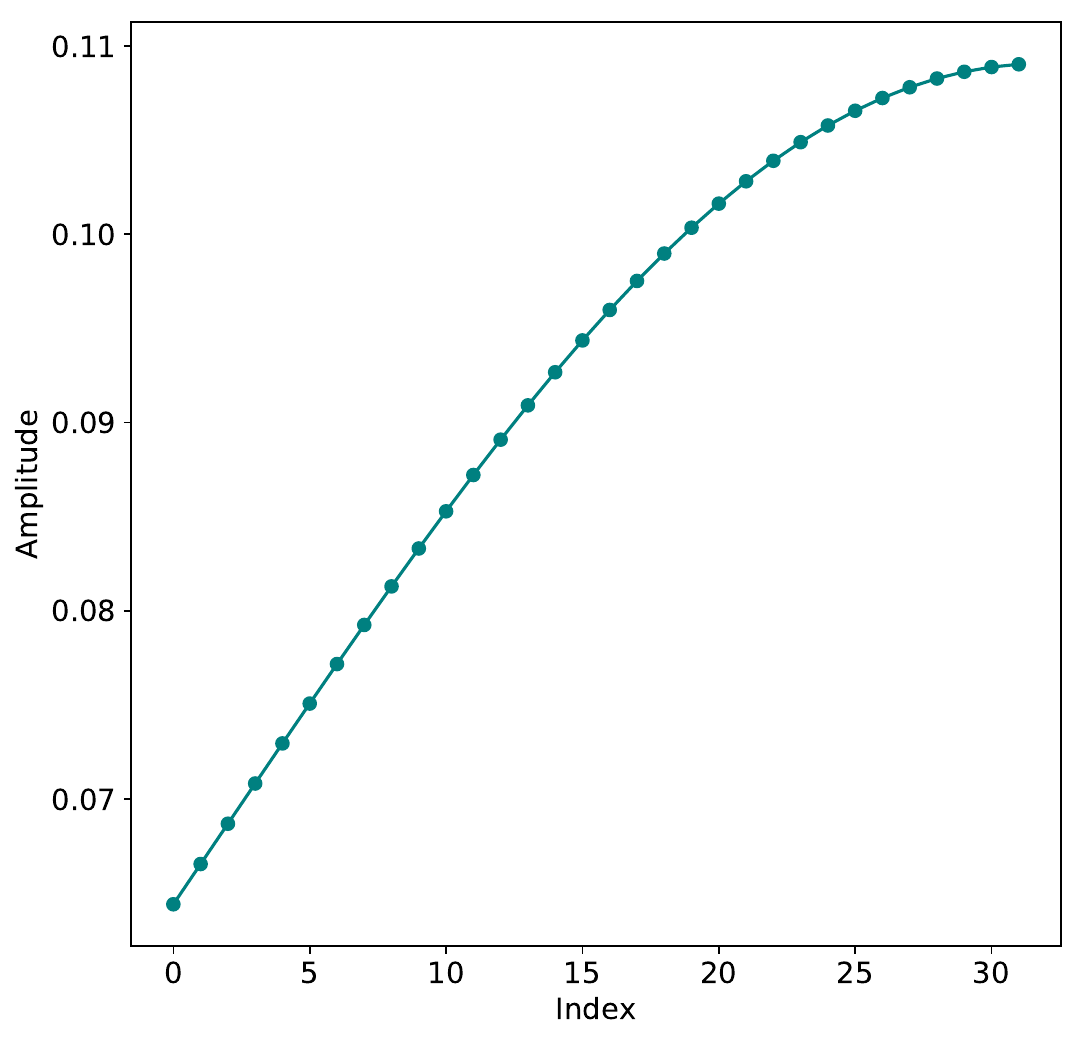}
      \caption{Eigenvector for lowest eigenvalue.}
      \label{fig:evec_001}
  \end{subfigure}
  \captionsetup{justification=centering}
  \begin{subfigure}[b]{0.40\textwidth}
      \centering
      \includegraphics[width=.9\textwidth]{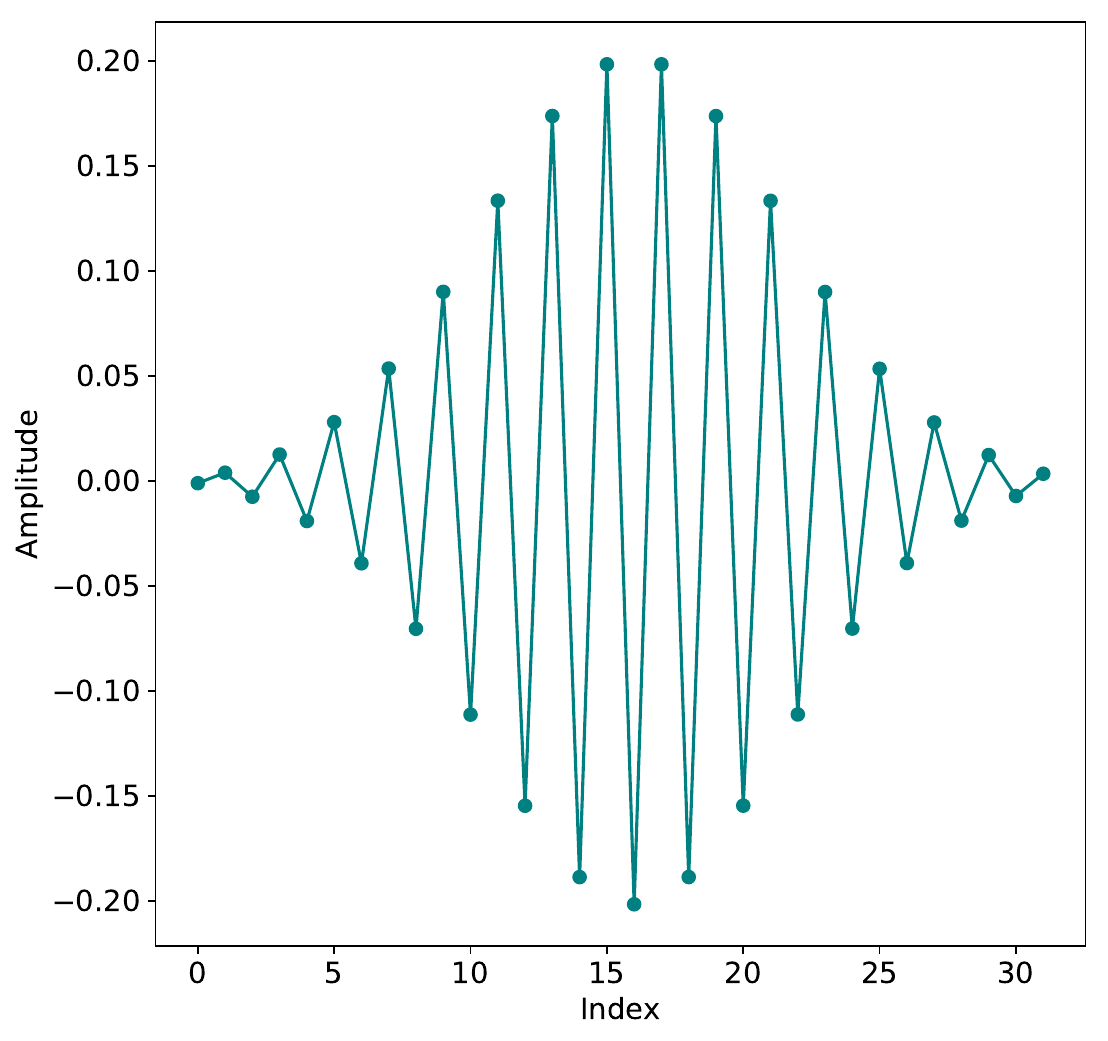}
      \caption{Eigenvector for largest eigenvalue.}
      \label{fig:evec_128}
  \end{subfigure}
  \caption{Comparison of lowest and largest eigenvectors for the coupled incompressible
  nozzle with 32 stations.}
  \label{fig:evecs}
\end{figure}

The eigenvectors corresponding to the lowest ($\lambda_0$) and highest ($\lambda_{N-1}$) eigenvalues are
shown in \Cref{fig:evecs} for a nozzle with 32 stations. 
Errors in the $1/y$ fit near $1/\kappa_s$ lead to an incorrect value
of $\alpha_0$. This applies a global shift to the solution that is detected
when the boundary conditions are applied for the next non-linear iteration.
This is beneficial as it means the condition number of $A$ does not need to be known 
exactly.
Conversely, if $\alpha_{N-1}$ is incorrect, a high frequency oscillation is 
added that is not detected by the boundary conditions. To remove this, the
non-linear solver effectively changes the sign of $\alpha_{N-1}$ on successive
iterations. This can be unstable and lead to divergence of the non-linear solver.

Referring to \Cref{fig:qsvt_vs_cheb}, 
the Cheb-LCU solver with  $\epsilon^{\rm sp}=0.01$ and $\epsilon^{\rm sp}=0.03$
polynomial fits are almost identical near $1/\kappa_s$,
but the latter has higher relative errors at around $y=0.2$ and $y=0.4$.
It is these, that lead to the factor of 2 difference in the number of 
iterations needed for the non-linear solver to converge.

% Implicit boundary conditions
% ----------------------------
\subsection{Handling boundary conditions}
\label{sec:bc_effect}

Classical linear equation solvers often adopt a \textit{matrix-free} approach where
in, say, a conjugate gradient solver, matrix-vector products are performed by calling
a function that loops over a graph rather than forming a matrix. These usually
apply the boundary conditions as a separate loop. This ensures that the solution of
the linear system satisfies the boundary conditions.

When forming the matrix to be passed to the quantum linear system solver, it is
usual to modify the matrix and RHS vector to implicitly incorporate the boundary
conditions. If the QLS accurately solves the system, the solution satisfies the
boundary conditions.
However, if the QLS only gives an approximate solution, this can lead to a degradation
in satisfying the boundary conditions as the non-linear iterations proceed.
This is easily remedied by explicitly imposing the boundary conditions at the
start of each non-linear iteration, but this can be overlooked if simply replacing 
the call to a classical solver with one to a quantum solver.

\begin{figure}[ht]
    \centering
    \includegraphics[width=0.45\linewidth]{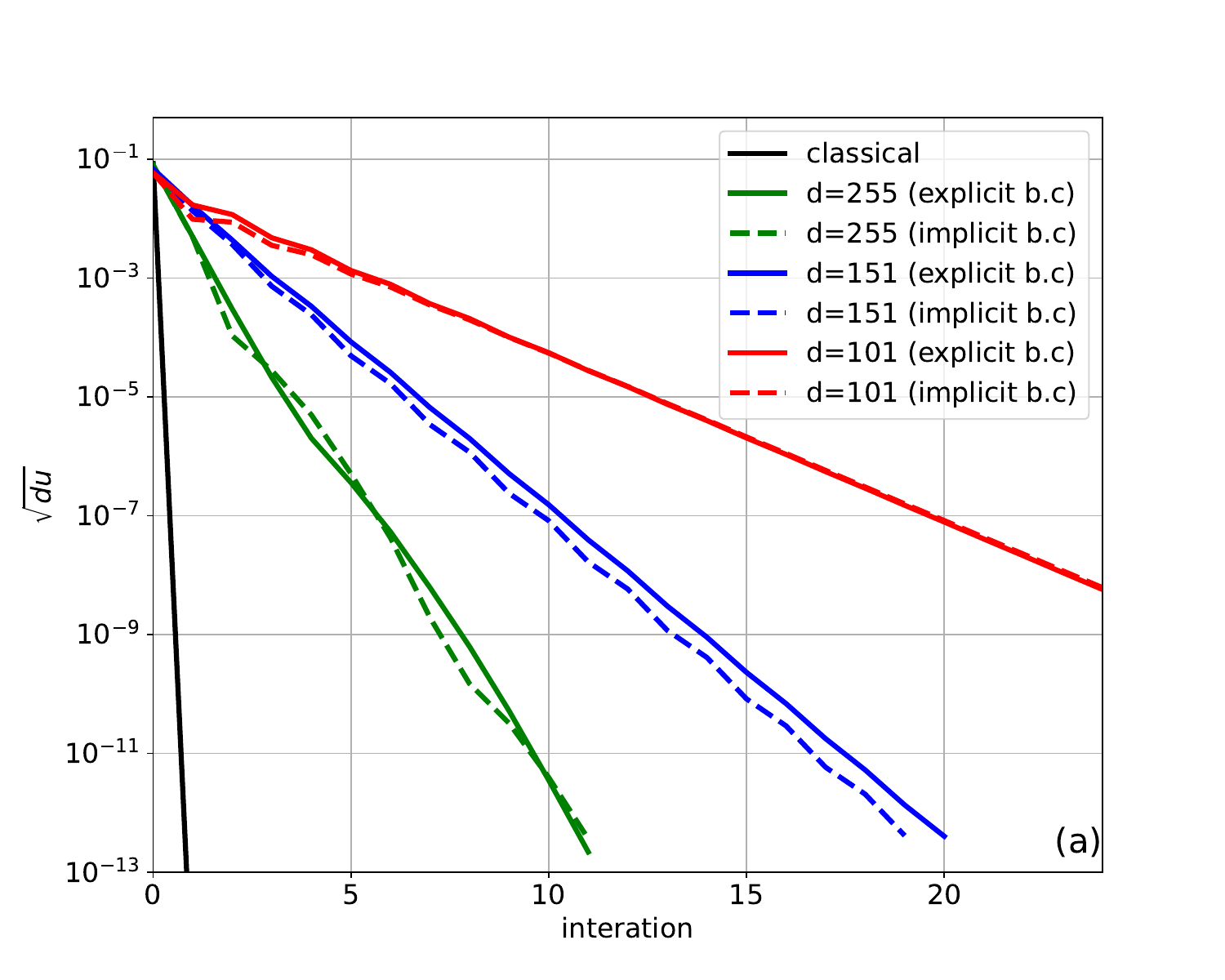}
    \includegraphics[width=0.45\linewidth]{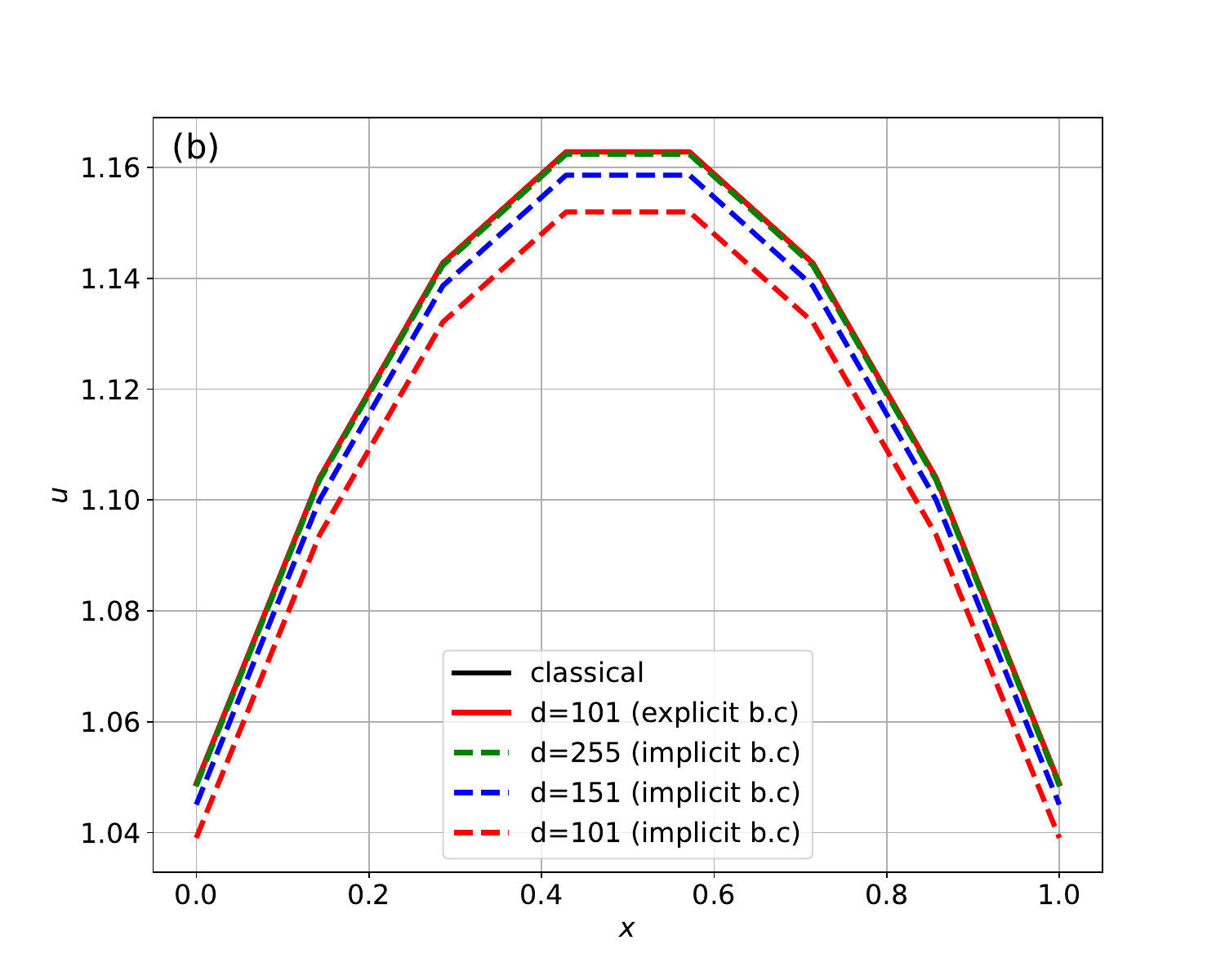}
    \caption{(a) Convergence histories for different polynomial degree (using QSVT), with explicit (solid colored lines) and implicit (dashed colored lines) boundary conditions. (b) The final velocity profile, here all the explicit boundary conditions cases converge to the classical profile, and thus we plot only the 101 degree example.}
    \label{fig:histories_bc}
\end{figure}

\Cref{fig:histories_bc} compares three QSVT solutions with varying
levels of approximation and with implicit and explicit boundary conditions.
The rate of convergence is largely independent of the boundary condition 
formulation.
All cases with explicit boundary conditions match the classical solution, as
does the implicit formulation with the accurate QSVT.
The other implicit QSVT solutions do not satisfy the boundary condition:
$u(0)=1.048$.
In more complicated calculations, a failure to satisfy the boundary conditions
can lead to divergence of the non-linear solver.

%% file: include/appendix-qls.tex
\section{Approximation of the inversion polynomial}
\label{app:chen_approx}

\subsection{Polynomial expansion with optimal relative error}
Various Chebyshev polynomial approximations for the inverse function $1/y$ can be found in the literature; see the summary in Ref.~\cite{Sunderhauf_etal2025}. The different
polynomial expansions differ in the type of error they optimize for a given degree, as well as in $\|P\|_{\max}$ and $\|\vec{a}\|_1$, which are
relevant to the efficiency of the QSVT and Cheb-LCU solvers; see Sec.~\ref{sec:qlss}. In this work, since we use a fixed degree, we adopt the polynomial 
approximation of Ref.~\cite{Gribling_etal2024}, which minimizes the relative error $\epsilon_{\rm rel}=\max_{y\in D_{\kappa_s}}|yP(y)-1|$ for the inverse function on 
$D_{\kappa_s} \equiv [-1,-1/\kappa_s]\cup[1/\kappa_s,1]$. The expansion is defined by the function
\begin{equation}
q_t(y)\equiv \frac{1-T_t\left(\frac{\kappa_s^{2}+1-2\kappa_s^{2}y^2}{\kappa_s^{2}-1}\right)/T_t\left(\frac{1+\kappa_s^{2}}{1-\kappa_s^{2}}\right)}{y},
\end{equation}
with $t=(d+1)/2$. The coefficients $\vec{a}$ in Eq.~\eqref{eq:f_inv} are obtained by a standard numerical procedure (we use the ``chebinterpolate'' method in NumPy). 
We can bound the corresponding relative error by 
\begin{align}
\epsilon_{\rm rel}
&=
\max_{y\in D_{\kappa_s}}
\left|
\frac{
T_t\left(\frac{\kappa_s^{2}+1-2\kappa_s^{2}y^2}{\kappa_s^{2}-1}\right)
}{
T_t\left(\frac{1+\kappa_s^{2}}{1-\kappa_s^{2}}\right)
}
\right|
\leq
\left|
\frac{1}{
T_t\left(\frac{1+\kappa_s^{2}}{1-\kappa_s^{2}}\right)
}
\right| \notag \\
&=
\frac{1}{2}
\left[
\left(\frac{\kappa_s + 1}{\kappa_s - 1}\right)^t
+
\left(\frac{\kappa_s + 1}{\kappa_s - 1}\right)^{-t}
\right]^{-1}.
\end{align}
where in the last equality we use the identity
$$
T_t(y) = \frac{1}{2}\left(\left(y-\sqrt{y^2-1}\right)^t+\left(x+\sqrt{x^2-1}\right)^t\right), \,\,\,\ \text{for } |y|>1.
$$

\subsection{Comparison to other expansions}

We find that other polynomial approximations perform worse in our CFD application. This is expected, since using relative error guarantees a good fit for the large eigenvalues,
which is essential for the convergence of the solution, as discussed in detail in Appendix~\ref{app-large-eval}.
In particular, we find that using the polynomial approximation that minimizes the uniform error $\epsilon_{\rm uni}=\max_{x\in D_\kappa}|P(y)-1/y|$ requires a
higher degree in order for the CFD to converge. This is because the approximation reaches negative values for $y>0$, resulting in divergent behavior.

\begin{figure}[ht]
    \centering
    \includegraphics[width=0.8\linewidth]{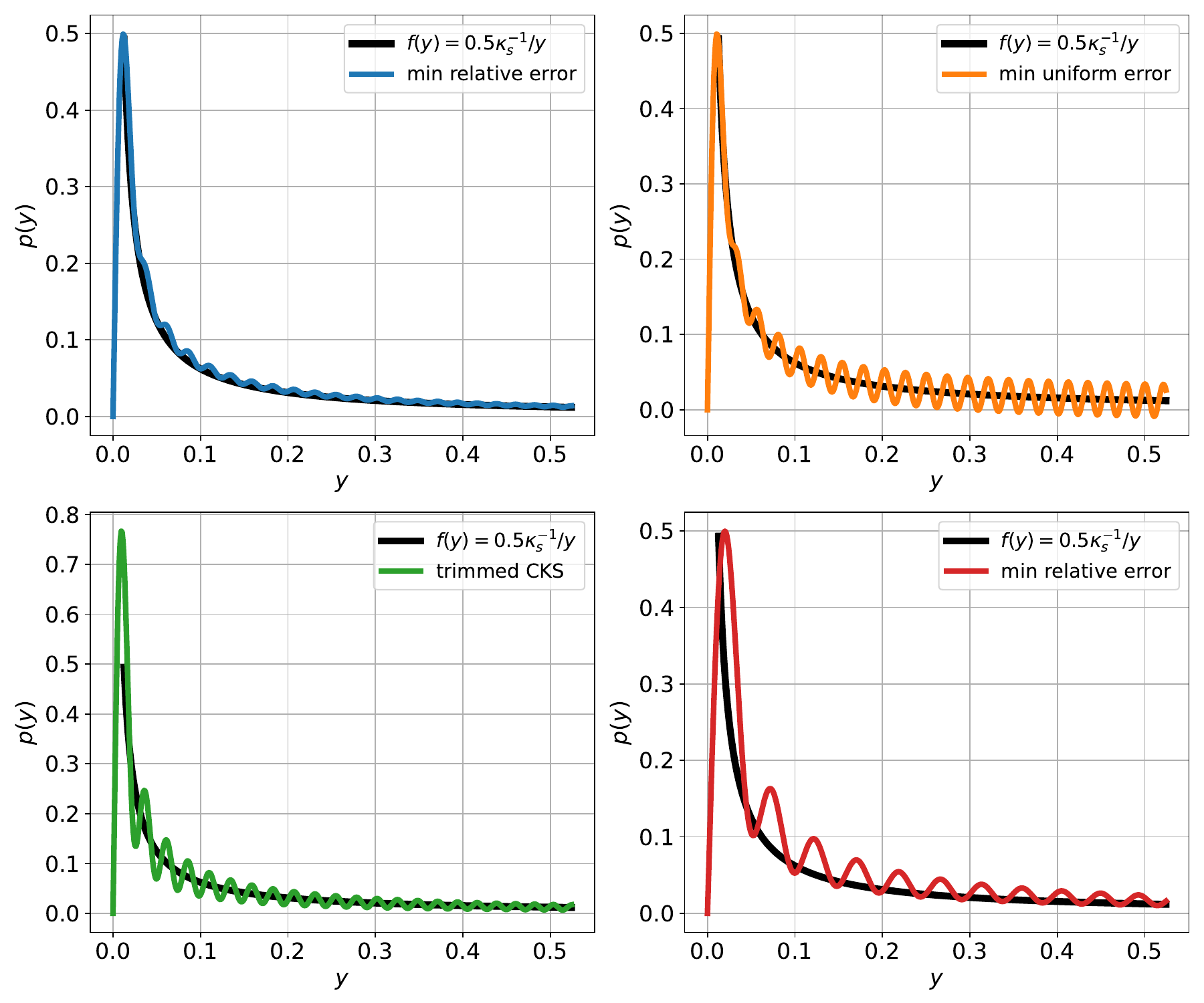}
    \caption{Different polynomial approximations for the inverse function. The upper left and lower right panels use a polynomial that 
    minimizes the relative error, for two different degrees, 255 and 127, respectively. This is the method used for our main results. The upper right panel shows the polynomial that
    optimizes the uniform error, whereas the lower left panel utilizes the approximation suggested in Ref.~\cite{Childs_etal2017}, trimmed to the given degree.
}
    \label{fig:different_cheb}
\end{figure}

In Fig.~\ref{fig:different_cheb} we compare the different Chebyshev expansion approaches for the inverse function. We can see that the approximation based
on the relative error gives meaningful results, remaining positive and following the target function, for degrees 255 and even 127. On the other hand, the polynomial that minimizes the uniform error has very large oscillations for the large eigenvalues and reaches negative values. We also include the polynomial defined in the original work
by Childs, Kothari, and Somma~\cite{Childs_etal2017} (CKS), trimmed to the requested degree. We can see that this approach overshoots the small eigenvalues and has moderate oscillations, which do not reach $p(y)<0$ for $y>0$. It is therefore expected to result in convergence of the CFD, but much more slowly than the polynomial with minimal relative error.

\subsection{Approximated expansion by approximated coefficients}
\label{app:approx_cheb_approx_coeffs}

\begin{figure}[ht]
    \centering
    \includegraphics[width=0.4\linewidth]{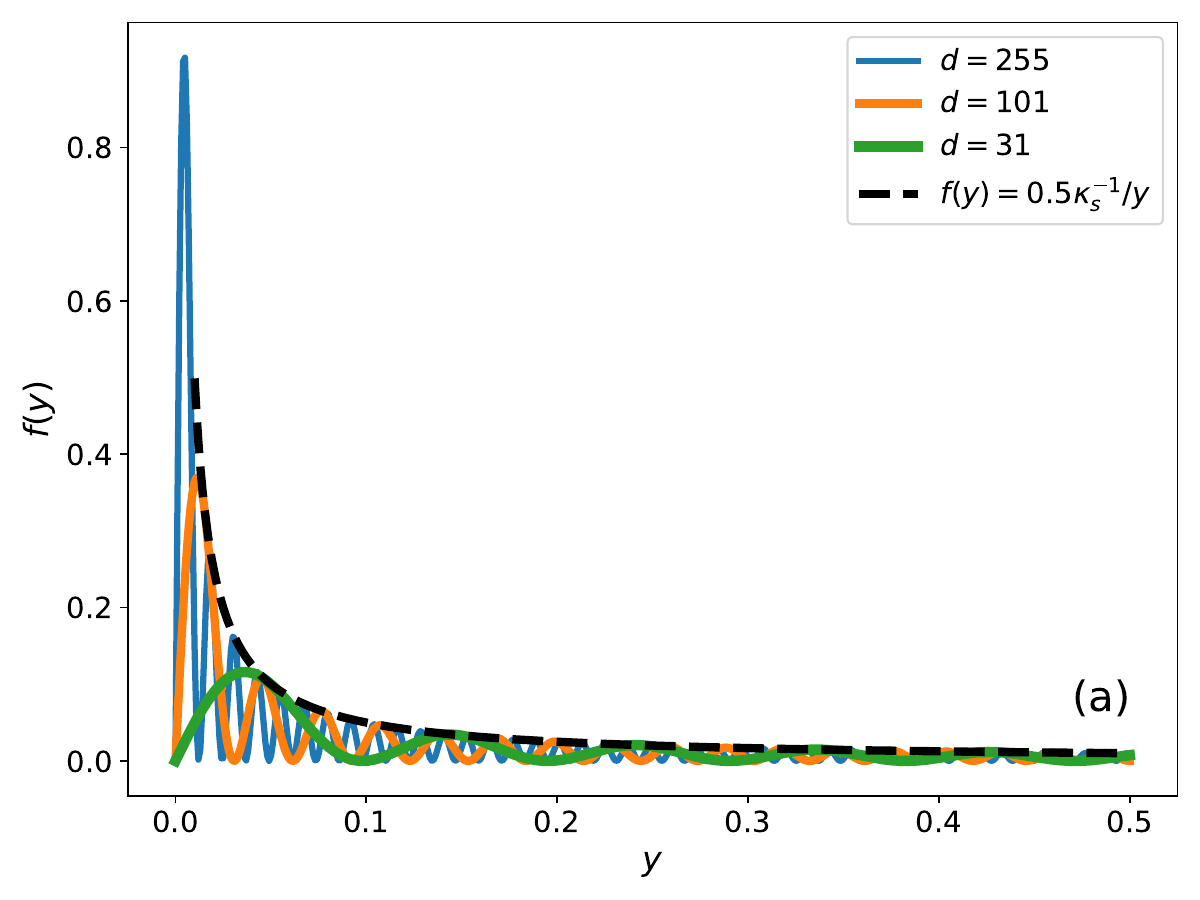}
    \includegraphics[width=0.4\linewidth]{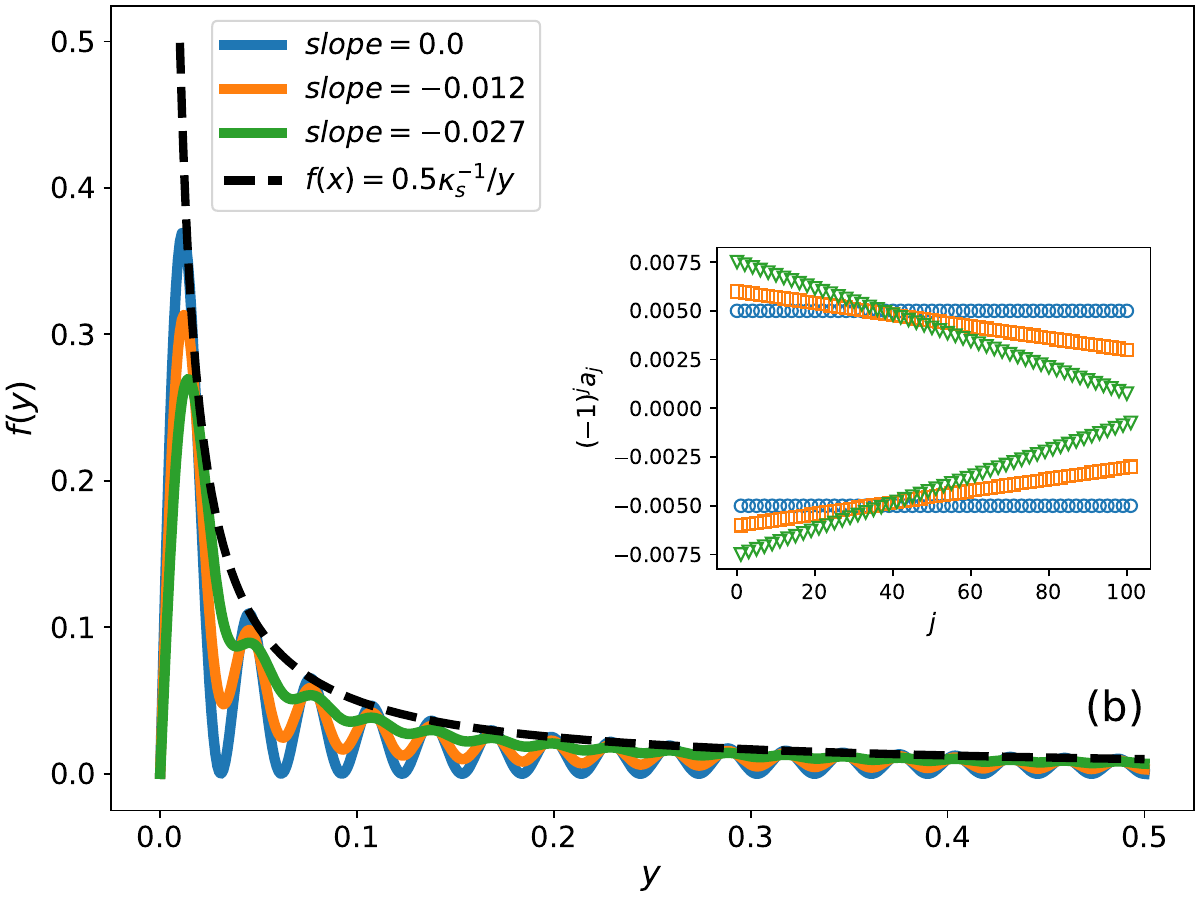}
    \caption{(a) The polynomial function with $\vec{a}$ a vector with constant entries, for different degree $d$. (b) The polynomial function with degree of 101,
    when taking a linear decreasing curve (presented in the inset) for the coefficients.}
    \label{fig:cheb_approx_coeff}
\end{figure}

In this section, we provide some technical details on the behavior of the Chebyshev coefficients $\vec{a}$ to approximate the inverse function.
In particular, we address the issue of why a linear decrease of the coefficients can give a sufficient approximation in general, and for our CFD application.
First, let us assume that $a_j=a$ is constant, for all $j$. One can show, that in this case the maxima (minima) of the polynomial in $y>0$ ($y<0$) follow the function $\propto 1/y$.
To see this, let us set constant coefficients in  the approximating polynomial in Eq.~\eqref{eq:f_inv}, and plug $y=\sin\theta$ for $\theta \in [-\pi/2,\pi/2]$:
$$
P(y)=a\sum^d_{j=0}(-1)^jT_{2j+1}\left(y\right)=a\sum^d_{j=0}(-1)^jT_{2j+1}\left(\sin\theta\right)=a\sum^d_{j=0}(-1)^jT_{2j+1}\left(\cos\left(\frac{\pi}{2}-\theta\right)\right).
$$
Then, recall that by definition, the Chebyshev polynomials satisfies,
$$
T_{2j+1}\left(\cos\phi\right) = \cos(n\phi),
$$
which brings us to
$$
P(y)=a\sum^{\frac{d-1}{2}}_{j=0}(-1)^j\cos\left((2j+1)\left(\frac{\pi}{2}-\theta\right)\right)=a\sum^{\frac{d-1}{2}}_{j=0}(-1)^j(-1)^j\sin\left((2j+1)\theta\right)=
a\sum^{\frac{d-1}{2}}_{j=0}\sin\left((2j+1)\theta\right).
$$
The final expression is a known trigonometric sum, which gives
$$
a\sum^{\frac{d-1}{2}}_{j=0}\sin\left((2j+1)\theta\right)=\frac{\sin^2((d+1)\theta)}{\sin\theta}.
$$
Thus, in summary, inserting back $y=\sin\theta$ we find that for $a_j=a$ to all $j$, the polynomial approximation of Eq.~\eqref{eq:f_inv} is
$$
P(y)=\frac{\sin^2((d+1)\arcsin(y))}{y},
$$
and its maxima for $y>0$ are at $a/y$. From the nominator of the result,
we can see that increasing the polynomial degree $d$ extends this relation for decreasing range in $y$, this is also illustrated in Fig.~\ref{fig:cheb_approx_coeff}a.

Now, if we perturb the constant coefficients expansion, we can expect that the resulting polynomial will still follow the $\propto 1/y$ trend. In particular,
if we take some linear decreasing function for $a_j$, we can see that the maxima at $y>0$ are slightly perturbed, while the minima points start to increase above 0. An example of this behavior is presented Fig.~\ref{fig:cheb_approx_coeff}(b).

\section{Block Encoding}
\label{app:be}
In this work block encoding is implemented using the banded-diagonal matrix encoding scheme of \cite{Lapworth&Sunderhauf2025}.
One can block-encode a matrix $D^{(i)}$, which contains only the diagonal $\vec{z}^{(i)}$, in two steps. First, load a matrix with the diagonal
elements on the main diagonal. For this we use a simple loading with a single qubit and $R_y$ rotations, uniformly controlled on the $|data\rangle$ variable, with the angles $\vec{\theta}^{(i)}=2\arcsin\left(\vec{z}^{(i)}/w^{(i)}\right)$ (padding the diagonal with zeros to obtain a full $2^n$ vector). In practice, we perform the 
uniform controls using the Graycode method according to Ref.~\cite{Mottonen_etal2004}. Second, we use an in-place adder to add $h^{(i)}$ to the $|data\rangle$ variable to shift the diagonal to its position in the original matrix $A$ (where positive values correspond to positions above the main diagonal and negative values to positions below the main diagonal). These two steps provide 
block encoding unitaries for the diagonal matrix, $U_{D^{(i)}}$. Finally, to block-encode $A$, we call a PREPARE-SELECT routine, where the PREPARE corresponds to loading the diagonals norms $\left(\sqrt{w^{(0)}},\sqrt{w^{(1)}},\dots,\sqrt{w^{(K-1)}}\right)/\sqrt{\sum^{K-1}_{i=0} w^{(i)}}$ on $k\equiv \lceil \log_2K\rceil$ qubits, 
and the SELECT operation selects a unitaries $U_{D^{(i)}}$.
\begin{figure}[ht]
    \centering
    \includegraphics[width=0.8\linewidth]{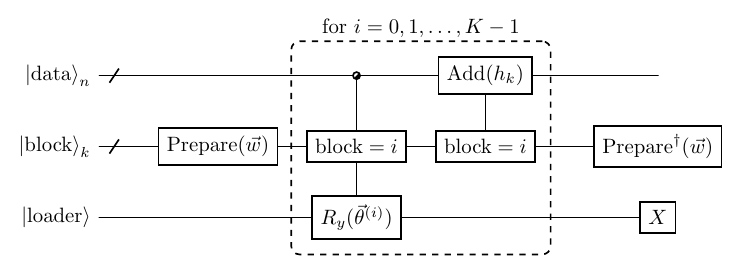}
    \caption{The model for block-encoding $A$ as a banded diagonals matrix. A PREPARE loads the 
    weights of the diagonals according to their max norm. Then a SELECT operation selects block-encoding a single diagonal matrix, using
    a diagonal loading with an extra qubit $|\text{loader}\rangle$ and a shift. The final $X$ operation on the $|\text{loader}\rangle$ qubit is just
    for setting the combined block zero variable $|\text{block}\rangle_k\otimes |\text{loader}\rangle$ to indicate the matrix $A$.
}
    \label{fig:be}
\end{figure}

\section{State Preparation}
\label{app:sp}
\begin{figure}[ht]
    \centering
    \includegraphics[width=0.8\linewidth]{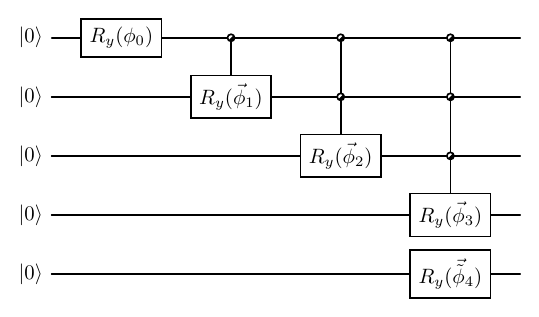}
    \caption{    Right: An approximated state preparation of arbitrary amplitudes, according to Ref.~\cite{Marin-Sanchez_etal2023}. Exact preparation corresponds to a cascade of uniform controls. Exchanging the last uniform control 
    with a single qubit rotation gives an approximated loading.}
    \label{fig:stateprep}
\end{figure}

The state-loading routine we use is based on the Grover-Rudolph technique~\cite{Grover&Rudolph2002}, which is constructed as a cascade of multiplex rotations: loading $2^l$ 
amplitudes is given by $2^k$ $k-$multi-controlled rotations for  $k=0,1,\dots, l-1$ qubits (where $k=0$ corresponds to uncontrolled rotation).
Then, each multiplex of size $k$ can be decomposed with a Gray-code into $2^k$ CX gates and  $2^k$ single qubit rotation~\cite{Mottonen_etal2004}. 
This results in a total of $2^l-1$ single-qubit rotations, i.e., $O(2^l)$ rotations, which gives the optimal asymptotic scaling for arbitrary state loading of $2^l$ amplitudes on $l$ qubits.
This state-loading design can be approximated by exchanging the last multiplexes by appropriate single qubit rotations, see Ref.~\cite{Marin-Sanchez_etal2023} and Fig.~\ref{fig:stateprep}.

